\documentclass[12pt]{article} 

\usepackage{amsmath}
\usepackage{graphicx}
\usepackage{array,epsfig,fancyheadings,rotating}
\usepackage{natbib}
\usepackage{url} 
\usepackage{amssymb}
\usepackage{epsfig}
\usepackage{stmaryrd}
\usepackage{amsfonts}
\usepackage{color}
\usepackage{float}
\usepackage{mathrsfs}
\usepackage{booktabs}
\usepackage{subfig}
\usepackage{multirow}
\usepackage{times}
\usepackage{bm}
\usepackage{xr} 
\usepackage{hyperref}
\usepackage[plain,noend]{algorithm2e}
\usepackage{enumerate}

\newcommand{\Y}{\mathbf{Y}}               
\newcommand{\X}{\mathbf{X}}               
\newcommand{\Z}{\mathbf{Z}}               
\newcommand{\V}{\mathbf{V}}
\newcommand{\U}{\mathbf{U}}
\newcommand{\A}{\mathbf{A}}
\newcommand{\B}{\mathbf{B}}
\newcommand{\C}{\mathbf{C}}

\newcommand{\I}{\mathbf{I}}  

\newcommand{\field}[1]{\mathbb{#1}}
\newcommand{\R}{\field{R}}

\newcommand{\E}{\field{E}}

\def\cov{{\mathrm{cov}}}
\def\var{{\mathrm{var}}}

\newtheorem{theorem}{Theorem}

\newtheorem{example}{Example}
\newtheorem{remark}{Remark}
\newtheorem{assumption}{Assumption}

\renewcommand{\theequation}{\thesection\arabic{equation}}

\begin{document}


\renewcommand{\baselinestretch}{2}

\markright{ \hbox{\footnotesize\rm Statistica Sinica
}\hfill\\[-13pt]
\hbox{\footnotesize\rm
}\hfill }

\markboth{\hfill{\footnotesize\rm CHUNG EUN LEE AND ZEDA LI} \hfill}
{\hfill {\footnotesize\rm MEAN INDEPENDENT COMPONENT ANALYSIS FOR MULTIVARIATE TIME SERIES} \hfill}

\renewcommand{\thefootnote}{}
$\ $\par


\fontsize{12}{14pt plus.8pt minus .6pt}\selectfont \vspace{0.8pc}
\centerline{\large\bf  MEAN INDEPENDENT COMPONENT ANALYSIS}
\vspace{2pt} 
\centerline{\large\bf FOR MULTIVARIATE TIME SERIES}
\vspace{.4cm} 
\centerline{Chung Eun Lee and Zeda Li} 
\vspace{.4cm} 
\centerline{\it  Paul H. Chook Department of Information Systems and Statistics, Baruch College}
 \vspace{.55cm} \fontsize{9}{11.5pt plus.8pt minus.6pt}\selectfont


\begin{quotation}
\noindent {\it Abstract:}
In this article, we introduce the mean independent component analysis for multivariate time series to reduce the parameter space. In particular, we seek for a contemporaneous linear transformation that detects univariate mean independent components so that each component can be modeled separately. The mean independent component analysis is flexible in the sense that no parametric model or distributional assumptions are made. We propose a unified framework to estimate the mean independent components from a data with a fixed dimension or a diverging dimension. We estimate the mean independent components by the martingale difference divergence so that the mean dependence across components and across time is minimized. The approach is extended to the group mean independent component analysis by imposing a group structure on the mean independent components. We further introduce a method to identify the group structure when it is unknown. The consistency of both proposed methods are established. Extensive simulations and a real data illustration for community mobility is provided to demonstrate the efficacy of our method. 

\vspace{9pt}
\par
\end{quotation}\par

\def\thefigure{\arabic{figure}}
\def\thetable{\arabic{table}}

\renewcommand{\theequation}{\thesection.\arabic{equation}}

\fontsize{12}{14pt plus.8pt minus .6pt}\selectfont

\newpage
\section{Introduction}\label{sec1}

Multivariate time series data is prevalent in a variety of fields, such as economics, business, engineering, climate science, and biomedicine. Often, high-dimensional time series data are encountered, where the modeling task  becomes more challenging since the modeling complexity increases dramatically with the dimension of the series. Due to this fact, the dimension reduction is essential and becomes the core of multivariate time series analysis. 

A primary concern in modeling a multivariate time series is searching for an elaborative but simplified underlying structure that substantially reduces the dimension of the parameter space. Many statistical methods have been proposed to reduce the dimension of the parameter space. On one hand, a factor model structure is assumed to achieve the dimension reduction on the data; see, for example, \citet{pena1987identifying,bai2002determining,forni2005generalized,pan2008modelling,lam2012factor} among others. On the other hands, there is a line of research imposing a simplified structure after applying a linear transformation to the data which includes the canonical correlation analysis \citep{box1977canonical}, the principal components analysis  \citep{stock2002forecasting}, the scalar component analysis \citep{tiao1989model}, the dynamic orthogonal component analysis \citep{matteson2011dynamic}. 
Furthermore, there is an extensive literature on the blind source separation 
in the signal processing. Such methods include the independent component analysis (ICA) \citep[among others]{back1997first,cardoso1998multidimensional,hyyarinen2001independent}, the second-order blind source separation \citep[among others]{tong1992finite, belouchrani1997blind,miettinen2016separation}. More recently, \citet{chang2018principal} have extended the principal components analysis (PCA) and sought for several lower-dimensional series that have no cross-correlations across different time lags. However, all the aforementioned approaches assume that the latent components are uncorrelated or independent, which is either highly simplified or extremely restrictive for multivariate time series. For a simple example, let the underlying component $\X_t=(x_{1,t}, \cdots, x_{p,t})^\top\in \R^p$ be a stationary process, where $x_{i,t}= \eta_{t-i+1}$ for $i=1, \cdots, p$. 
On one hand, the latent independent components do not exist unless the sequence $\{\eta_{t}, \eta_{t-1}, \cdots, \eta_{t-p+1}\}$ is independent. This independence condition seems to be a strong constraint for a time series to be satisfied.
On the other hand, the latent components are uncorrelated if the sequence $\{\eta_{t}\}$ is generated from an uncorrelated white noise such as the all-pass ARMA(1,1) model, $\eta_{t}=\phi\eta_{t-1}+u_t-\phi^{-1}u_{t-1}$ with a zero mean, i.i.d. sequence of $\{u_{t}\}$, and $|\phi|<1$.
However, it is obvious that there exists a dependence beyond correlation among $x_{i,t}= \eta_{t-i+1},~ i=1, \cdots, p$. Thus, imposing uncorrelateness on latent components may overlook this important dynamics and highly simplify the dependence of the components, which may not be optimal. 
To this end, it would be ideal to search for the latent componets that satisfy more constrained assumption than uncorrelatedness but less restrictive than independence.   

In this article, we propose an alternative approach that 
searches for novel latent components which better align with the modeling and forecasting in practice, since the conditional mean is considered. This differentiates our method from most of the existing methods.  
We focus on seeking the components that are mutually mean independent across time so that each component can be modeled separately by univariate time series models, and reduce the number of parameters in modeling. More specifically, our goal is to estimate the orthogonal matrix $\A$ that decomposes the data $\Y_t \in \R^{p}$ into the mean independent components, $\X_t=(x_{1,t}, \cdots x_{p,t})^\top$ such that \vspace{-0.5cm}
\[\A^\top \Y_t = \X_t,~~\E\left(x_{i,t}\mid \mathcal{F}_{t}^{(-i)}\right) = \E(x_{i,t})~\text{almost surely for }i=1, \cdots, p,\vspace{-0.5cm}
\]
where $\mathcal{F}_{t}^{(-i)}=\sigma\left\{\X_{t}^{(-i)}, \X_{t-1}^{(-i)},\cdots\right\}$ is the current and past information of $\X_{t}^{(-i)}\in \R^{p-1}$ which contains all the elements of $\X_t$ except $x_{i,t}$, and the components of $\X_t$ are mutually mean independent across time. Thus $\{x_{i,t-1}, x_{i,t-2}, \cdots\}$ are sufficient to model the behavior of $x_{i,t}$ for each $i$ and supports to build a separate univariate time series model with the least amount of loss on the conditional mean of $x_{i,t}$. This is especially important because the conditional mean is the optimal predictor in the mean squared error sense. 
Hence, compared to the existing approaches, our mean independent component analysis can produce more accurate forecasting. Inspired by the method in \citet{chang2018principal}, we further extend our approach by imposing a group structure that segments the time series into several mean independent lower-dimensional subseries and propose a method and an algorithm to estimate the unknown group structure. 
Since the mean independence is the relationship that lies between the uncorrelatedness and the independence, our approaches naturally bridge the gap between the PCA approaches and the ICA approaches. Furthermore, no parametric model or distributional assumptions are imposed on the conditional mean $\E(x_{i,t}\mid x_{i,t-1}, x_{i,t-2}, \cdots)$, thus our approaches are model-free in this aspect. Moreover, our approaches can detect both linear and nonlinear dependence. Consequently they can be widely applicable to real word data exhibiting nonlinear dependence including the economical data such as interest rates, exchange rates, economic price data \citep{tsay1998testing, terasvirta2010modelling, wang2022testing}, the environmental data such as temperature, precipitation, and sea ice dynamics, \citep{munch2023constraining, marwan2021nonlinear}, and the physiological signals such as gait variability and heart rate variability \citep{calderon2023revisiting, li2023robust}.  
Our approaches are built upon the metric called martingale difference divergence (MDD) of \citet{ShaoZhang2014JASA} that perfectly captures the mean dependence between two vectors. To the best of our knowledge, the theoretical investigation of MDD for the large dimension has been mostly developed under the i.i.d. data. However, there are relatively fewer studies of MDD for the large dimension under the dependent data. In this article, we establish the asymptotic results of our methods under both fixed dimension and diverging dimension, which seems to be new to the literature and requires totally different techniques. 

The rest of the article is organized as follows. We provide a brief review of the martingale difference divergence and the martingale difference correlation in Section~\ref{sec:review}. We introduce our mean independent component analysis in Section~\ref{sec:MIC} and introduce the group mean independent component analysis in Section~\ref{sec:GMIC} with the theoretical investigations. All the numerical results are summarized in Section~\ref{sec:simulation}. In Section~\ref{sec:real data}, we present a real data application and Section~\ref{sec:conclusion} concludes. All the proofs and additional theoretical results, simulation results, real data application results, and figures are gathered in the supplementary material.

\section{Review of Martingale Difference Divergence}\label{sec:review}
To introduce our new approaches, we briefly review the martingale difference divergence (MDD) and its scale invariant metric, the martingale difference correlation (MDC), in \cite{ShaoZhang2014JASA} and \cite{ParkShaoYao2015EJS}. For $\U \in \R^{r}$ and  $\V\in \R^q$, where $r$ and $q$ are positive integers, the metrics MDD and MDC measure the mean dependence between two vectors $\V$ and $\U$, i.e.,
$\E(\V\mid \U)=\E(\V) ~\mbox{almost surely}.$
More specifically, if $\E(\|\V\|^2+\|\U\|^2)< \infty$ where $\|\cdot\|$ is the Euclidean norm, then
	${\text{MDD}}^2(\V\mid \U)=-\E\left[\left\{\V-\E(\V)\right\}^\top\left\{\V^{'}-\E(\V^{'})\right\}\|\U-\U^{'}\|\right] \in \R$,
${\text{MDC}}^2(\V\mid \U)=\frac{{\text{MDD}}^2(\V\mid \U)}{\left\{\text{tr}\left({\text{var}}(\V)^2\right)\right\}^{1/2}\left\{{\text{dvar}}(\U)\right\}^{1/2}}\in [0,1]$,
where $(\V^{'}, \U^{'})$ is an independent copy of $(\V, \U)$ and $\text{dvar}(\U)$ is the distance variance of $\U$ in \cite{szekely2007measuring}.
The key property of MDD and MDC is that \vspace{-0.5cm}
\begin{eqnarray}\label{eq:MDD_prop}
\text{MDD}^2(\V\mid \U)=0, ~\text{MDC}^2(\V\mid \U)=0 \text{ if and only if } \E(\V\mid \U)=\E(\V) ~\mbox{almost surely},\end{eqnarray}\vspace{-1.5cm}

\noindent thus they fully characterize the mean independence of $\V$ on $\U$.
For given random vectors $(\U_i,  \V_i)_{i=1}^{n}$ from the joint distribution of $(\U,\V)$, the sample $\text{MDD}^2$ and the sample $\text{MDC}^2$ are defined as 
$\widehat{{\text{MDD}}}^2(\V \mid \U)= \frac{-1}{n^2}\sum_{i,j=1}^{n} (\V_i -\overline{\V})^\top (\V_j-\overline{\V})\|\U_i-\U_{j}\|$, $\widehat{\text{MDC}}^2(\V\mid \U) = \frac{\widehat{{\text{MDD}}}^2(\V \mid \U)}{\left\{tr\left(\widehat{\text{var}}(\V)^2\right)\right\}^{1/2}\left\{\widehat{\text{dvar}}(\U)\right\}^{1/2}}$,
where $\overline{\V}=\frac{1}{n}\sum_{k=1}^{n} \V_{k}$, $\text{tr}\left(\widehat{\text{var}}(\V)^2\right)=\text{tr}\left(\frac{1}{n}\sum_{k=1}^{n} \left(\V_{k}-\overline{\V}\right)\left(\V_{k}-\overline{\V}\right)^\top\right)^2$ $= \text{tr}\left(\frac{1}{n^2}\sum_{i,j=1}^n A_{i,j}^2\right)$, $A_{i,j}=a_{i,j} - a_{i,\cdot} - a_{\cdot, j}+a_{\cdot, \cdot}$, $a_{i,j}=\frac{1}{2}\|\V_i-\V_j\|^2$, $a_{i,\cdot}=\frac{1}{n}\sum_{j=1}^n a_{i,j}$, $a_{\cdot,j}=\frac{1}{n}\sum_{i=1}^n a_{i,j}$, $a_{\cdot, \cdot}=\frac{1}{n^2}\sum_{i,j=1}^n a_{i,j}$, $\widehat{\text{dvar}}(\U)=\frac{1}{n^2}\sum_{i,j=1}^n B_{i,j}^2$, $B_{i,j}=b_{i,j} - b_{i,\cdot} - b_{\cdot, j}+b_{\cdot, \cdot}$, $b_{i,j}=\|\U_i-\U_j\|$, and $b_{i,\cdot}$, $b_{\cdot,j}$, $b_{\cdot, \cdot}$ are defined similarly to $a_{i,\cdot}$, $a_{\cdot,j}$ and $a_{\cdot,\cdot}$. 

\section{Mean Independent Component Analysis}\label{sec:MIC}
\subsection{Mean Independent Component Model}

We shall first briefly review the existing methods, the dynamic orthogonal compoments in \citet{matteson2011dynamic} and the second-order blind source separation in \citet{belouchrani1997blind}, which are closely related to our approach. The existing approaches assume that the data 
$\Y_t = \B \Z_t$, where $\E(\Y_t) = \E(\Z_t)={\bf 0}$, $\var(\Y_t) = \var(\Z_t)=\I$, $\B=(\B_{\cdot 1}, \cdots, \B_{\cdot p})$ is an orthogonal matrix, $\Z_t = (z_{1,t}, \cdots, z_{p,t})^\top \in \R^p$ are the underlying components that are uncorrelated to each other, i.e., $\cov\left(z_{i,t}, z_{j,t-h}\right) = 0~\text{for }i \neq j, ~\text{all lags }h=0, 1, 2,\cdots$.

Next, we consider distinct components and introduce an alternative approach. In particular, 
\vspace{-1.2cm}
\begin{eqnarray}\label{eq:single_model}
\Y_t = \A \X_t, 
\end{eqnarray}\vspace{-1.5cm}

\noindent where $\E(\Y_t) = \E(\X_t)={\bf 0}$, $\var(\Y_t) = \var(\X_t)=\I$, $\A=(\A_{\cdot 1}, \cdots, \A_{\cdot p})$ is an orthogonal matrix, $\X_t = (x_{1,t}, \cdots, x_{p,t})^\top$ are the mean independent components, i.e., $\E\left\{x_{i,t}\mid \mathcal{F}_{t}^{(-i)}\right\} = \E(x_{i,t})~\text{almost surely}$, $\mathcal{F}_{t}^{(-i)}=\sigma\left\{\X_{t}^{(-i)}, \X_{t-1}^{(-i)},\cdots\right\}$, $\X_{t}^{(-i)} = (x_{1,t}, \cdots,x_{i-1,t}, x_{i+1,t},\cdots,x_{p,t})^\top$. 

\noindent Similar to the other existing methods, the orthogonal matrix $\A$ and the mean independent components $\X_t$ are not unique in terms of signed permutation. In other words, $\Y_t = \A \mathbf{P}_{\pm}^\top \left(\mathbf{P}_{\pm}\X_t\right)$ produces the same data $\Y_t$, where $\mathbf{P}_{\pm}$ is a signed permutation matrix. However, in modeling or forecasting perspectives, discovering the mean independent components up to a signed permutation is sufficient. Therefore, we seek to identify the mean independent components up to a signed premutation; see Section 2.4 in \citet{matteson2011dynamic}. 
\begin{remark}\label{remark:SMIC1} \rm{We shall make several remarks on the comparison between our method and the existing methods.
It is worth mentioning that our approach is essentially different from the dynamic orthogonal components in \citet{matteson2011dynamic} and the second-order blind source separation in \citet{belouchrani1997blind}. All approaches aim to reduce the number of parameters involved in the modeling. Therefore, all methods seek for the orthogonal matrix $\A$ or $\B$ so that each component can be equipped with a separate univariate model. However, \citet{matteson2011dynamic} and \citet{belouchrani1997blind} search for serially uncorrelated components by adopting covariances. 
Unless the data is Gaussian or solely generated from linear time series models, separating components through the linear metric is not sufficient. If the forecasting or modeling is the primary concern, we can alternatively separate the components by examining the contributions toward the conditional mean of each components, and seek for serially mean independent components $\X_t$. Since the conditional mean is the best predictor in terms of the mean squared error, we shall lose less prediction accuracy by focusing on the mean independent components. Also, it is worth pointing out that our approach summarizes both the linear and nonlinear dependence of components based on their contribution toward the conditional mean. Hence, our approach can be more robust and flexible to the dependence and the distribution of the data. Due to these facts, our approach can be viewed as a nonlinear analog of the dynamic orthogonal components in \citet{matteson2011dynamic} or the second-order blind source separation in \citet{belouchrani1997blind}. Following the existing approaches, we assume $\var(\Y_t) = \var(\X_t)=\I$. However, it is possible to extend our model representation to the data $\Y_t$ with $\var(\Y_t) \neq \I$. Then we lose the orthogonality constraint on $\mathbf{A}$ which affects the computational time since optimizing the objective function with the orthogonality constraint is easier than optimizing the objective function without a constraint. In terms of the implementation, our approach fully measures the mean dependence through $\text{MDD}$ to estimate the hidden components which is a moment-based method. Thus, comparing to the existing methods, our approach is also simple to implement and can be widely applicable in practice since no parametric or distributional assumptions are imposed; see Section~\ref{subsec:estimation_SMIC} for the objective function.
}
\end{remark}

\subsection{Estimation of Mean Independent Components}\label{subsec:estimation_SMIC}
Our specific goal is to seek $\A$ that decomposes our data $\Y_t$ into mean independent components $\X_t$. 
Notice that the mean independence between $x_{i,t}$ and $x_{j,t-h}$, $h \geq 0$ can be fully captured by $\text{MDD}$ due to its property in (\ref{eq:MDD_prop}), i.e., 
$\text{MDD}^2(x_{i,t}\mid x_{j,t-h}) =0, ~i \neq j,~h=0, 1, \cdots$.
Based on our model in (\ref{eq:single_model}), it implies that $\text{MDD}^2(\A_{\cdot i}^\top \Y_{t}\mid \A_{\cdot j}^\top \Y_{t-h}) =0, ~i \neq j,~h=0, 1, \cdots$.
Motivated by this fact, we propose to search $\A$ by minimizing the objective
function, \vspace{-0.7cm} \begin{eqnarray}\label{eq:single_optimize_pop}
\mathcal{S}_{h_0}(\widetilde{\A}) &=&\left(\frac{1}{\text{tr}\{\widetilde{\A}\widetilde{\A}^\top\}}\right)\sum_{h=0}^{h_0}\sum_{i=1}^p\sum_{j \neq i}\text{MDD}^2(\widetilde{\A}_{\cdot i}^\top \Y_{t}\mid \widetilde{\A}_{\cdot j}^\top \Y_{t-h}), 
\end{eqnarray}\vspace{-1cm}

\noindent where $\mathcal{S}_{h_0}(\widetilde{\A}) \geq 0$ and it becomes zero at the true $\A$, i.e., $\mathcal{S}_{h_0}({\A})=0$.
Hence, our estimator $\widehat{\A}$ is 
\vspace{-1cm}
\begin{eqnarray}\label{eq:single_optimize}\widehat{\A} = \text{argmin}_{\widetilde{\A}^\top \widetilde{\A}= \widetilde{\A}\widetilde{\A}^\top =\I}\widehat{\mathcal{S}}_{h_0}(\widetilde{\A}),
\end{eqnarray}\vspace{-1.0cm}

\noindent where $\widehat{\mathcal{S}}_{h_0}(\widetilde{\A})=\left(\frac{1}{tr\left\{\widetilde{\A}\widetilde{\A}^\top\right\}}\right)\sum_{h=0}^{h_0}\sum_{i=1}^p \sum_{j \neq i}\widehat{\text{MDD}}^2(\widetilde{\A}_{\cdot i}^\top \Y_{t} \mid \widetilde{\A}_{\cdot j}^\top \Y_{t-h})$.
\begin{remark}\label{remark:SMIC2}{\rm We shall make several remarks. We note that the denominator of the objective function in (\ref{eq:single_optimize_pop}) does not affect the optimization procedure. Thus, in practice, the denominator has less impact on the estimation. However, the denominator is required to show the theoretical properties when $p$ diverges. More specifically, the denominator is necessary in order to obtain Lipschitz continuity of $\mathcal{S}_{h_0}$ with respect to the distance in (\ref{eq:D-distance_SMIC}); see Lemma~1 in the supplementary material. 
Furthermore, selecting $h_0$ has been a common question in the literature and proposing a method to select $h_0$ is beyond the scope of this paper. As mentioned in the literature, generally, relatively small $h_0$ is used since major
dependence is often at the short time lag for stationary time series and more noises can be added if large $h_0$ is selected; see \citet{lam2012factor} and \citet{wang2019factor}. Lastly, it is possible to provide weights for difference lags and construct the objective function similar to \citet{matteson2011dynamic}. However, we use our objective function in (\ref{eq:single_optimize}) considering the close connection with the method in \citet{chang2018principal} which is described in Section~\ref{sec:GMIC}. Also, we observe that our approach with the equal weights in (\ref{eq:single_optimize}) provides encouraging performance under our simulation study; see Section~\ref{sec:simulation}.  
}
\end{remark}\vspace{-0.5cm}

To optimize the objective function in (\ref{eq:single_optimize}), we adopt the parameterization estimation approach for $\widetilde{\A}$ in \citet{matteson2011dynamic} and \citet{belouchrani1997blind}. 
Specifically, we set $\widetilde{\A} = \prod_{1 \leq i<j\leq p} \Gamma_{ij}(\theta_{ij})$, where $\Gamma_{ij}(\theta_{ij})$ is an identity matrix with $(i,i)$th and $(j,j)$th elements replaced by $\text{cos}(\theta_{ij})$, and $(i,j)$th and $(j,i)$th elements replaced by $\text{sin}(\theta_{ij})$ and $-\text{sin}(\theta_{ij})$, respectively. 
Note that our objective function is locally convex. By following the common practice, we try multiple starting points to avoid local minima. Specifically,  we employ Latin hypercube sampling to uniformly sample the space of $\theta_{ij}$ to generate 200 multiple initial values. We select the best one in terms of our objective function to begin the process. When $p$ is large, we suggest estimating the mean independent components sequentially so that the optimization is performed with reasonable computational time. However, it is important to note that estimation uncertainty accumulates at each stage of the sequential process compared to the joint estimation. Here, we have a tradeoff between computational complexity and statistical efficiency. For more details regarding estimating the mean independent components sequentially, we refer readers to \cite{matteson2017independent} in section 2.4.




Next, we shall introduce the asymptotic results for our estimation procedure for both fixed and diverging $p$. 
Since $\A$ is not identifiable in terms of signed permutation, we adopt the following distance from \cite{fan2008modelling} which measures the closeness of two orthogonal matrices and establish the consistency of our estimator. After arranging the orders of the columns based on the similarity for $\B=(\B_{\cdot 1}, \cdots, \B_{\cdot p}) \in \R^{p \times p}$ and $\C=(\C_{\cdot 1}, \cdots, \C_{\cdot p}) \in \R^{p \times p}$, the $\mathcal{D}$-distance is defined as \vspace{-0.5cm}
\begin{eqnarray}\label{eq:D-distance_SMIC}\mathcal{D}(\B, {\C}) = \left(p-\sum_{i=1}^p  \left| \B_{\cdot i}^\top {\C}_{\cdot i}\right|\right)^{1/2}.
\end{eqnarray}
Notice that for any orthogonal matrices $\B$ and $\C$, the distance is nonnegative, i.e.,  $\mathcal{D}(\B, \C)\geq 0$, and smaller $\mathcal{D}$ implies that two orthogonal matrices $\B$ and $\C$ are more similar. We make the following assumptions, under which we derive the consistency of our estimator. 

\begin{assumption}\label{assum:1}Let $\mathcal{H}_{\mathcal{D}}$ be the space that consists of orthogonal matrices. 
\begin{enumerate}
\item There exists an orthogonal matrix $\A \in \mathcal{H}_{\mathcal{D}}$ that minimizes the objective function $\mathcal{S}_{h_0} (\cdot)$. Furthermore, the minimum value of $\mathcal{S}_{h_0} (\cdot)$ is obtained at an orthogonal matrix $\widetilde{\A}$ if and only if $\mathcal{D}(\A,\widetilde{\A})=0$. 
\item When $p$ is fixed, the data $(\Y_t)_{t=1}^n$ is strictly stationary and $\beta$-mixing process. There exists $\delta>0$ such that $\E\|\Y_t\|^{6+3\delta}<\infty$ and the $\beta$-mixing coefficients $\beta(k)=O(k^{-(2+\delta')/\delta'})$ for $\delta' \in (0, \delta)$. 
\item When $p$ diverges with $n$, the set $\mathcal{H}_{\mathcal{D}}$ is the union of $\mathcal{H}_{p}$, where $\mathcal{H}_{p}$ is the set containing all $p\times p$ orthogonal matrices.
\item When $p$ diverges, the data $(\Y_t)_{t=1}^n$ is strictly stationary and $\beta$-mixing process. There exists $\delta>0$ such that $\E\left\|\frac{1}{\sqrt{p}}\Y_{t}\right\|^{6+6\delta}<C$ for all $p$, where $C$ is a finite constant. The $\beta$-mixing coefficients satisfy $
\sum_{k=1}^{\infty} k \beta(k)^{\delta/(1+\delta)}<\infty$.
\end{enumerate}
\end{assumption}

The first condition in Assumption~\ref{assum:1} ensures that $\A$ is the unique minimizer of $\mathcal{S}_{h_0}$ in the sense of $\mathcal{D}$. This condition is necessary to show the theoretical properties, and it is assumed in \citet{matteson2011dynamic}. The second and fourth conditions are used to bound certain moments when $p$ is fixed or $p$ diverges, where similar $\beta$-mixing condition in the fourth condition is also used in time series literature; see \citet{hjellvik1996linearity} and \citet{dette2004some}. The third condition is assumed so that $\widehat{\A}$ is well defined and measurable when $p$ diverges and guarantees the consistency of $\widehat{\A}$ in terms of $\mathcal{D}$ if $\widehat{\mathcal{S}}_{h_0}(\widetilde{\A})$ converges uniformly to $\mathcal{S}_{h_0}(\widetilde{\A})$; see the proof of Theorem~\ref{thm:fixed_large_p_SMIC} in the supplementary material.  

\begin{theorem}\label{thm:fixed_large_p_SMIC} 
\begin{enumerate} 
\item Let $h_0$ and $p$ be fixed integers.
 Under the first and second conditions in Assumption~\ref{assum:1}, we have $\mathcal{D}(\A, \widehat{\A}) \rightarrow^p 0~\text{as }n\rightarrow \infty$. 
\item Let $h_0$ be a fixed integer and $p$ grow with $n$ such that $p\sqrt{p}/\sqrt{n} \rightarrow 0$ as $n\rightarrow \infty$.
 Under the first, third, and fourth conditions in Assumption~\ref{assum:1}, we have $\mathcal{D}(\A, \widehat{\A}) \rightarrow^p 0~\text{as }n\rightarrow \infty$. 
\end{enumerate}
\end{theorem}
Theorem~\ref{thm:fixed_large_p_SMIC} indicates that when the dimension $p$ is fixed or $p$ grows with $n$ such that $p\sqrt{p}/\sqrt{n} \rightarrow 0$, our estimator gets closer to the true $\A$ in terms of the $\mathcal{D}$-distance as sample size increases. 
It is worth mentioning that the assumptions are made towards the data $\Y_t$ but not on the mean independent components $\X_t$ to study the asymptotic properties. It seems hard to obtain the convergence rate for $\mathcal{D}(\mathbf{A}, \widehat{\mathbf{A}})$ under the current assumptions since the main theorems we used only guarantee $\mathcal{D}(\mathbf{A}, \widehat{\mathbf{A}})\rightarrow 0$. However,  it is possible to extend the theorem so that our approach is theoretically justified under a weaker assumption for $p$ by following the proof in \citet{chang2018principal},  e.g.,  $p(\log(p))^{4/\gamma'}/n\rightarrow 0$,  $\gamma'>0$.  This is reported in our supplementary material. Also, in the literature, often a parametric model or distributional assumptions are made to derive the theoretical results. 
However, our approach does not assume a parametric model or a distribution of the data. In terms of the theoretical investigations, the existing methods studied asymptotic properties for fixed $p$ while our approach examines both fixed $p$ and growing $p$. Moreover, the theoretical properites of MDD is mostly investigated for i.i.d. data. However, we derive the asymptotic results of MDD with dependent data, which is relatively less studied in the literature. Notice that our objective function depends on $n$ when $p$ diverges. Thus, showing the consistency is far from trivial and requires new techniques. 

\section{Group Mean Independent Component Analysis}\label{sec:GMIC}
\subsection{Group Mean Independent Component Model}
In many applications, the observed data might be produced from components that have a certain group structure, resulting in lower-dimensional subseries. Inspired by the work of the PCA for time series (TS-PCA) in \cite{chang2018principal}, we extend the mean independent component analysis with a group structure the so-called group mean independent component analysis. More precisely, we consider 
\vspace{-0.4cm}
\begin{eqnarray}\label{eq:group_model}
\Y_t = \A \X_t, 
\end{eqnarray}\vspace{-1.4cm}

\noindent where $\E(\Y_t ) = \E(\X_t)={\bf 0}$, $\var(\Y_t) = \var(\X_t)=\I$, $\A=(\A_{ 1}, \cdots, \A_{ m})$ is an orthogonal matrix, $\A_{ i} \in \R^{p \times p_i}$, $\sum_{i=1}^m p_i = p$, $\X_t = \left\{(\X_{t}^{(1)})^\top, \cdots, (\X_{t}^{(m)})^\top\right\}^\top\in \R^p$ is the group mean independent components, $\X_{t}^{(i)} = \left(x_{1,t}^{(i)}, x_{2,t}^{(i)}, \cdots, x_{p_i,t}^{(i)}\right)^\top\in \R^{p_i}$, i.e., \vspace{-0.3cm}
\begin{eqnarray}\label{eq:group_MIC}
\E\left\{\X_{t}^{(i)}\mid \mathcal{F}_{t}^{(-i)}\right\} = \E(\X_{t}^{(i)})~\text{almost surely},
\end{eqnarray}\vspace{-1.1cm}

\noindent here, $\mathcal{F}_{t}^{(-i)}=\sigma\left\{\X_{t}^{(-i)}, \X_{t-1}^{(-i)},\cdots\right\}$, $\X_{t}^{(-i)} = \left\{(\X_{t}^{(1)})^\top, \cdots,(\X_{t}^{(i-1)})^\top, (\X_{t}^{(i+1)})^\top,\cdots,(\X_{t}^{(m)})^\top\right\}^\top\in \R^{p-p_i}$. As mentioned in \cite{chang2018principal}, the orthogonal matrix $\A$ is not identifiable but $\{\text{span}(\A_i)\}_{i=1}^m$ are unique; see Section 2.1 in \cite{chang2018principal}. Thus, our goal is to estimate the identifiable spaces $\{\text{span}(\A_i)\}_{i=1}^m$.  
\begin{remark}\label{remark:1}{\rm We shall make a few remarks. While TS-PCA in \cite{chang2018principal} segments elements of $\X_t$
by the uncorrelatedness, our approach divides the groups of $\X_t$ by the mean independence in (\ref{eq:group_MIC}).
Because of this fact, our method has several distinct features. In particular, since the TS-PCA
separates the groups through the autocovariance matrices, it suffers from the same limitations mentioned in Remark~\ref{remark:SMIC1}. Thus, our approach can be more robust and flexible to different forms of
dependence and different distributions of the data. Furthermore, as mentioned in \cite{chang2018principal}, TS-PCA cannot separate uncorrelated components if different blocks share the same eigenvalues of their linear matrix. 
However, our approach bypasses the limitation of the existing method since our approach directly estimates the group mean independent components by optimizing an objective function which considers the mean dependence between two groups. Also, if there is prior knowledge of the group, our approach can easily incorporate the known group structure. Lastly, the correlatedness implies
the mean dependence, which suggests a nested structure for a particular $i$th group, i.e., $\text{span}(\overline{\A}_{ i}) \subseteq\text{span}(\A_{ i}), i \in \{1,2,\cdots, m\}$, after the same alignments for $\A$ and $\overline{\A}$, where $\overline{\A} = ( \overline{\A}_{1}, \cdots ,\overline{\A}_{ l})$ is the orthogonal matrix that generates the
block components that are uncorrelated in TS-PCA. Due to this fact, for a given $p=\text{dim}(\Y_t)$, it is possible for our approach to have a different number of groups compared to TS-PCA. More specifically, it is possible that the TS-PCA selects more groups than our approach, i.e., $m \leq l$.  
To better understand the differences, we provide one example. \begin{example} \rm{Suppose the data $\Y_t$ is generated by two groups of mean independent components, i.e., $\Y_t = \A \X_t = \A \begin{pmatrix}\X_{t}^{(1)}\\ \X_t^{(2)}\end{pmatrix}$,
where one group $\X_t^{(1)}=(x_{1,t}^{(1)}, \cdots, x_{p_1,t}^{(1)})^\top$ has components which are mutually uncorrelated but mean dependent across time and the other group $\X_t^{(2)}=(x_{1,t}^{(2)}, \cdots, x_{p-p_1,t}^{(2)})^\top$ has components that are correlated and mean dependent. 
}
\end{example}
\noindent Under the model assumption in TS-PCA, the true group structure becomes $(1,1, \cdots, 1, p-p_1)$ which indicates that $\{x_{i,t}^{(1)}\},~i=1, \cdots, p_1$ construct $p_1$ different groups while $\X_t^{(2)}$ form a single group. Thus, there are $l=p_1+1$ different groups for TS-PCA. On the other hand, under our framework, there are $m=2$ different groups ($\X_t^{(1)}$, $\X_t^{(2)}$) and the true group structure is $(p_1, p-p_1)$. 
Hence, it is possible to have different number of groups and different group structures for the existing method and our approach. 
}
\end{remark}

\subsection{Estimation of Group Mean Independent Components}\label{subsec:estimation_GMIC}
Suppose the group structure $(p_1, \cdots, p_m)$ are known, and $m$ is fixed. Similar to the mean independent components in Section~\ref{sec:MIC}, we have $\text{MDD}^2\left(\X_{t}^{(i)} \mid \X_{t-h}^{(j)}\right)=\sum_{k=1}^{p_i}\text{MDD}^2(x_{k,t}^{(i)} \mid \X_{t-h}^{(j)})=0, ~i \neq j, ~h=0,1,\cdots,$ due to (\ref{eq:group_MIC}) and this is equivalent to $\text{MDD}^2(\A_{ i}^\top \Y_{t} \mid \A_{ j}^\top \Y_{t-h})=\sum_{k=1}^{p_i}\text{MDD}^2\left((\A_{i})_{\cdot k}^\top \Y_{t} \mid \A_{ j}^\top \Y_{t-h}\right)=0, ~i \neq j, ~h=0,1,\cdots,$ where $(\A_{i})_{\cdot k}$ is the $k$th column in $\A_i$.
By this fact, we define the objective function as \vspace{-0.3cm}
\begin{eqnarray}\label{eq:obj_group}
\mathcal{G}_{h_0}(\widetilde{\A}) 
&=&\left(\frac{1}{tr\left(\widetilde{\A}\widetilde{\A}^\top\right)}\right)^{1/2}\sum_{h=0}^{h_0}\sum_{i=1}^{m} \sum_{k=1}^{p_i}\sum_{j \neq i} {\text{MDD}}^2\left((\widetilde{\A}_{ i})_{\cdot k}^\top \Y_{t} \mid \widetilde{\A}_{j}^\top \Y_{t-h}\right),
\end{eqnarray}\vspace{-1cm}

\noindent where $\mathcal{G}_{h_0}(\widetilde{\A})\geq 0$ and it becomes zero at the true orthogonal matrix $\A$ with the true group structure, i.e., $\mathcal{G}_{h_0}({\A})=0$. Notice that the denominator is different from the one in $\mathcal{S}_{h_0}$. This is mainly due to the fact that there are fewer elements that construct $\mathcal{G}_{h_0}$ and we use a different distance to measure the accuracy of our approach. In other words, while there are $h_0p(p-1)$ different $\text{MDD}^2$ added in $\mathcal{S}_{h_0}$, there are $h_0p (m-1)$ different $\text{MDD}^2$ that construct $\mathcal{G}_{h_0}$. Also, adopting a different distance is inevitable to measure the accuracy of the group mean independent component analysis since the spaces $\{\text{span}(\A_i)\}_{i=1}^m$ are identifiable but not the signed permuted $\A$. Therefore, a different distance is necessary to measure the closeness between two spaces; see (\ref{eq:D-distance_GMIC}) for the distance.

For the estimation of $\A$, we shall utilize\\ $\widehat{\mathcal{G}}_{h_0}(\widetilde{\A})= \left(\frac{1}{tr\left(\widetilde{\A}\widetilde{\A}^\top\right)}\right)^{1/2}\sum_{h=0}^{h_0}\sum_{i=1}^{m} \sum_{j \neq i} \widehat{\text{MDD}}^2\left(\widetilde{\A}_{ i}^\top \Y_{t} \mid \widetilde{\A}_{ j}^\top \Y_{t-h}\right)$,
and search for $\A$ \vspace{-0.5cm}
\begin{eqnarray}\label{eq:group_optimize}
\widehat{\A}=\text{argmin}_{\widetilde{\A}^\top \widetilde{\A} = \widetilde{\A} \widetilde{\A} ^\top =\I} \widehat{\mathcal{G}}_{h_0}(\widetilde{\A}).
\end{eqnarray}\vspace{-1.5cm}

\noindent Since $\{\text{span}(\A_{i})\}_{i=1}^m$ are identifiable, we shall establish the consistency for our estimator ${\A}$ with a similar $\widetilde{\mathcal{D}}$-distance from \citet{chang2018principal}, i.e., After arranging the orders of $\{\B_i\}_{i=1}^m$ and $\{\C_i\}_{i=1}^m$ based on the similarity, \vspace{-0.3cm}
\begin{eqnarray}\label{eq:D-distance_GMIC}\widetilde{\mathcal{D}}(\B, \C) =\max_{i=1, \cdots, m} \widetilde{\mathcal{D}}(\B_i, \C_i)= \max_{i=1, \cdots, m}\left(p_i-tr\left( \B_{ i} \B_{i}^\top \C_{i}\C_{i}^\top \right)\right)^{1/2},
\end{eqnarray}\vspace{-1.3cm}

\noindent where $\B_i,\C_i\in \R^{p \times p_i}$ are semi-orthogonal matrices, $\widetilde{\mathcal{D}}^2(\B_i, \C_i)=p_i-tr\left( \B_{ i} \B_{ i}^\top \C_{i}\C_{i}^\top\right)$. Smaller $\widetilde{\mathcal{D}}$-distance indicates that $\{\text{span}(\B_i)\}_{i=1}^m$ and $\{\text{span}(\C_i)\}_{i=1}^m$ are closer. 

\begin{theorem}\label{thm:fixed_large_p_GMIC} \begin{enumerate} \item Let $h_0 $, $m$, and $p$ be fixed integers. Assume that $(p_1, \cdots, p_m)$ are known.
 Under the first and second conditions in Assumption~\ref{assum:1} with respect to $\mathcal{G}_{h_0}$, we have $\widetilde{\mathcal{D}}(\A, \widehat{\A}) \rightarrow^p 0~\text{as }n\rightarrow \infty$. 
\item  Let $h_0$ and $m$ be fixed integers. Assume that $(p_1, \cdots, p_m)$ are known. Let $p$ grow with $n$ such that $p/\sqrt{n} \rightarrow 0$ as $n\rightarrow \infty$. Under the first, third, and fourth conditions in Assumption~\ref{assum:1} with respect to $\mathcal{G}_{h_0}$, we have $\widetilde{\mathcal{D}}(\A, \widehat{\A}) \rightarrow^p 0~\text{as }n\rightarrow \infty$. 
\end{enumerate}
\end{theorem}

We note that the mean independent components correspond to the special case of the group mean independent components where $m=p$ and each component constructs a separate group. However, the growing rate for $p$ in Theorem~2 is different from the one in Theorem~1. The main reason is that we obtain different convergence rates for $\left| \widehat{\mathcal{S}}_{h_0}(\widetilde{\mathbf{A}})-\mathcal{S}_{h_0}(\widetilde{\mathbf{A}})\right|$ and $\left| \widehat{\mathcal{G}}_{h_0}(\widetilde{\mathbf{A}})-\mathcal{G}_{h_0}(\widetilde{\mathbf{A}})\right|$ under the assumption that the group number $m$ is fixed. This eventually leads to two distinct growing rates of $p$ for two approaches.  Further,  as we assume $m$ is fixed when $p$ diverges,  this implies that $p_j/p\rightarrow c$ for $j=1, \cdots, m$ where $0\leq c\leq 1$ is a constant.  It would be interesting to extend the theoretical justification for $m\rightarrow \infty$, this seems to be quite challenging. First, when $m \rightarrow \infty$, it is not guaranteed that our distance $\widetilde{\mathcal{D}}(\mathbf{A} , \widehat{\mathbf{A}})$ is well defined since we have infinite $\{\widetilde{\mathcal{D}}(\mathbf{A}_i , \widehat{\mathbf{A}}_i)\}_{i=1}^{\infty}$. Moreover, Lemma 4 in the supplementary material will not hold for the objective function $\mathcal{G}_{h_0}$ under the case $m \rightarrow \infty$. Due to these facts 
showing the consistency result for $m\rightarrow \infty$ will require new techniques and we shall leave this for future work.


In practice, the true group structure $(p_1, \cdots, p_m)$ may not be given and needs to be estimated. We shall follow the maximum cross correlation approach in \cite{chang2018principal}, which is stemmed from the ratio-based estimator in the factor model literature. We first select $r$ which is the number of two pairs of components that have at least one time lag showing a significant mean dependence. By searching for the pairs that have significant serial mean dependence, we can estimate $(p_1, \cdots, p_m)$. More specifically, we estimate $r$ by \vspace{-0.3cm} 
\begin{eqnarray}\label{eq:group_info}\widehat{r} = \text{argmax}_{1\leq j \leq c_0p_0} \frac{\widehat{\text{M}}_j}{\widehat{\text{M}}_{j+1}}, 
\end{eqnarray}\vspace{-1.3cm}

\noindent where $c_0=0.75$ which is recommended in \citet{chang2018principal}, $p_0=p(p-1)/2$ is the total number of two pairs of $p$ components, $\widehat{\text{M}}_1 \geq \widehat{\text{M}}_2 \geq \cdots \geq \widehat{\text{M}}_{p_0}$ are every pairs of $\widehat{\text{M}}(i, j)$ in the descending order, \vspace{-0.3cm}
\begin{eqnarray}\label{eq:M}\widehat{\text{M}}(i, j) = \max_{0\leq h\leq h_0}\left\{ \widehat{\text{MDC}}^2(\widehat{x}_{i,t}\mid \widehat{x}_{j,t-h}), \widehat{\text{MDC}}^2(\widehat{x}_{j,t}\mid \widehat{x}_{i,t-h})\right\},
\end{eqnarray}\vspace{-1.3cm}

\noindent and $\widehat{\X}_t=\widehat{\A}^\top \Y_t=(\widehat{x}_{1,t}, \cdots, \widehat{x}_{p,t})^\top$. 
 \noindent Once $r$ is estimated, we define an undirected graph $G=(V,E)$, where $V=\{1,2,\cdots, p\}$ is the vertex set and $E$ is the set of edges such that $e_{i,j}=e_{j,i}=1$ if $\widehat{\text{M}}(i,j) \in (\widehat{\text{M}}_1, \cdots, \widehat{\text{M}}_{\widehat{r}})$ or $e_{i,j}=e_{j,i}=0$ if $\widehat{\text{M}}(i,j) \not\in (\widehat{\text{M}}_1, \cdots, \widehat{\text{M}}_{\widehat{r}})$. Two components $\widehat{x}_{i,t}$ and $\widehat{x}_{j,t}$ belong to the same group if the vertices $i$ and $j$ are directly or indirectly connected, i.e., $e_{i,j}=1$ or there exists $\{v_1, v_2, \cdots, v_w\} \subset  V~\text{such that}~ e_{i,v_1} = e_{v_1, v_2}= \cdots = e_{v_w,j} = 1$. We further summarize the entire process to estimate our group mean independent components when the group structure is not provided in Algorithm \ref{alg:group}.

  \RestyleAlgo{ruled}
{\begin{algorithm}[!h]
	\caption{\small Algorithm to estimate the group mean independent components and the group structure}\label{alg:group}
	\KwData{\small $\Y_t = (y_{1,t}, \dots, y_{p,t})^\top$}
	\KwResult{\small $\widehat{\A}=(\widehat{\A}_{1}, \cdots, \widehat{\A}_{m})$, $\widehat{\X}_t=(\widehat{x}_{1,t}, \cdots \widehat{x}_{p,t})^\top$, and $(\widehat{p}_1,\cdots, \widehat{p}_{\widehat{m}})$.}
	{\bf Step 1}: \small  Begin the algorithm by setting $m=p$ and $p_1=\cdots=p_m=1$. Obtain an initial estimate $\widehat{\A}^{(0)}$ by minimizing $\widehat{\mathcal{S}}_{h_0}(\cdot)$ in \eqref{eq:single_optimize} and estimate the components by $\widehat{\X}^{(0)}_t = (\widehat{\A}^{(0)})^\top \Y_t$.\\
	{\bf Step 2}: Compute $\widehat{\text{M}}(i,j)$ defined in \eqref{eq:M} for every two pairs of components and arrange $\widehat{\text{M}}(i,j)$ in the descending order.\\
{\bf Step 3}: Select ${r}$ through the ratio-based estimator in \eqref{eq:group_info} with presepcified $c_0$ and collect $(\widehat{\text{M}}_1, \cdots, \widehat{\text{M}}_{\widehat{r}})$, where $\widehat{\text{M}}_k$ is the $k$th largest $\widehat{\text{M}}(i,j)$.\\
	{\bf Step 4}: Based on the collected $(\widehat{\text{M}}_1, \cdots, \widehat{\text{M}}_{\widehat{r}})$, create an undirected graph $G=(V,E)$, where $V=\{1,2,\cdots, p\}$ is the vertex set and $E$ is the set of edges such that $e_{i,j}=e_{j,i}=1$ if $\widehat{\text{M}}(i,j) \in (\widehat{\text{M}}_1, \cdots, \widehat{\text{M}}_{\widehat{r}})$ or $e_{i,j}=e_{j,i}=0$ if $\widehat{\text{M}}(i,j) \not\in (\widehat{\text{M}}_1, \cdots, \widehat{\text{M}}_{\widehat{r}})$. \\
	{\bf Step 5}: Based on the graph in Step 4, estimate the group structure, $(\widehat{p}_1, \cdots, \widehat{p}_{\widehat{m}})$. For instance, two components $i$ and $j$ belong to the same group if the vertices $i$ and $j$ are directly connected or indirectly connected, i.e., $e_{i,j}=1$ or there exists $\{v_1, v_2, \cdots, v_w\} \subset  V~\text{such that}~ e_{i,v_1} = e_{v_1, v_2}= \cdots = e_{v_w,j} = 1.$\\ 
{\bf Step 6}: Estimate $\widehat{\A}^{(1)}$ by minimizing $\widehat{\mathcal{G}}_{h_0}(\cdot)$ in (\ref{eq:group_optimize}) with $(\widehat{p}_1, \cdots, \widehat{p}_{\widehat{m}})$ obtained from Step 5. Permute $\widehat{\A}^{(1)}$ based on the estimated group structure and estimate the components by $\widehat{\X}^{(1)}_t = (\widehat{\A}^{(1)})^\top \Y_t$, where $\widehat{\A}^{(1)}$ is the permuted matrix. \\
{\bf Step 7}: Repeat Step 2 - Step 6 until the estimated group structure, $(\widehat{p}_1, \cdots, \widehat{p}_{\widehat{m}})$, does not change and $\|\widehat{\A}^{(i+1)}-\widehat{\A}^{(i)}\|_F<\epsilon$, where $\widehat{\A}^{(i+1)}$ and $\widehat{\A}^{(i)}$ are the estimates of $\A$ after $i$th and $(i-1)$th iterations, and $\epsilon$ is the prespecified tolerance.  
\end{algorithm}}
\vspace{-0.3cm}
\begin{remark}\rm{
We shall make several remarks on Algorithm \ref{alg:group} and our group mean independent component analysis. If the group information is known, the group structure is directly incorporated into the objective function (\ref{eq:group_optimize}) by pre-specifying the number of groups $m$ and $p_1, \cdots, p_m$. We suggest estimating semi-orthogonal matrix $\mathbf{A}_i$, $i=1,\cdots,m$, sequentially using the optimization solver with orthogonality constraint proposed by \cite{wen2013feasible}, which has an efficient computational time; see \cite{wen2013feasible} for more details. If the group structure is unknown, we suggest using Algorithm \ref{alg:group}. Step 2 - Step 5 estimates the group structure and Step 6 estimates $\A$. We use our mean independent components $\widehat{\X}_t^{(0)}$ in Section~\ref{sec:MIC} as an initial estimate for the group mean independent components. The intention is to group different components into one group that still show the mean dependence in $\widehat{\X}_t^{(0)}$, where components are as mutually mean independent as possible. 
This strategy is often used in the literature; see \citet{stogbauer2004least}, \citet{cardoso1998multidimensional}, \citet{gruber2009hierarchical} among others. 
Furthermore, it is desirable to conduct the false discovery rate-based multiple testing with MDD to assess the mean independence for every two pairs of components. However, constructing a test is very nontrivial since we only have $\widehat{\X}_t$ and the true $\X_t$ is unknown, thus there will be an estimation effect. Also, each test shall involve a bootstrap procedure which requires intensive computations. In this sense, the ratio-based approach in Algorithm \ref{alg:group} is more computationally efficient to estimate the group structure. We further note that there is a new method for the estimation of $r$; see the variant of ratio-based estimator discussed in \citet{chang2015high} and \citet{han2024simultaneous}. We use the ratio-based estimator in (\ref{eq:group_info}) by following the practice of the implementation in \citet{chang2018principal}. Algorithm \ref{alg:group} generally requests several iterations to converge and shows an improvement after each iteration. For instance, the algorithm stops after 2 or 3 iterations under our simulation examples. Besides, the estimated $\widehat{\A}^{(1)}$ and $(\widehat{p}_1, \cdots, \widehat{p}_{\widehat{m}})$ obtained through Step 1 - Step 6 generate reasonable results in our simulation study and they are not greatly different from the final estimates after a few iterations until the algorithm converges. 
Lastly, the theory to show the consistency of our estimator of the group structure is very challenging since it involves showing the consistency of $\widehat{r}$ and $(\widehat{p}_1, \cdots, \widehat{p}_{\widehat{m}})$. As mentioned in the existing literature, showing the consistency of the estimator $\widehat{r}$ is not straightforward because it is hard to show that the maximum of the ratio between $\widehat{\text{M}}_j$ and $\widehat{\text{M}}_{j+1}$ is occured when $j=r$. Even if there is a new approach that estimates $r$ with the consistency, the challenge still remains for showing the consistency of $(\widehat{p}_1, \cdots, \widehat{p}_{\widehat{m}})$.  To obtain this consistency, it is necessary to show $P(\widehat{e}_{i,j} \neq e_{i,j}| \widehat{r} = r)\rightarrow 0$ for any $i,j$, which is difficult and we cannot follow the similar arguments in \citet{chang2018principal} and \citet{han2024simultaneous}. Notice that the estimation procedure of our approach and the one in \citet{chang2018principal} are fundamentally different. While our approach optimizes an objective function to search $\mathbf{A}$, the method in \citet{chang2018principal} applies eigen decomposition to their linear matrix. Thus, $\widehat{\mathbf{M}}(i,j)$ in (\ref{eq:M}) has more complicated form which cannot be expressed by the multiplication of a particular matrix and vectors. Thus, $\left|\widehat{\mathbf{M}}(i,j)-{\mathbf{M}}(i,j)\right|$ cannot be controlled by the difference between a matrix and its sample counterpart. Also, unlike the methods in \citet{chang2018principal} and \citet{han2024simultaneous}, it is hard to obtain the convergence rate of $\|\widehat{\mathbf{A}} - \mathbf{A}\|_2$, $\|\cdot\|_2$ is a spectral norm, for our approach which makes the theoretical analysis even more challenging. We shall leave this for future work.
}
\end{remark}




\section{Numerical Simulations}\label{sec:simulation}

\subsection{Mean Independent Component Analysis}\label{subsec:sim_SMIC}
We examine the finite sample performance of our approach for mean independent component analysis (MICA) in Section~\ref{sec:MIC} and consider two examples with various distributions. In particular, the stationary data is generated by $\Y_t = \A \X_t$, where the orthogonal matrix $\A$ is generated through \texttt{randortho} function in \texttt{pracma} R package by following \citet{stewart1980efficient}, and $\X_t = (x_{1,t}, \cdots, x_{p,t})^\top$ is generated by \vspace{-0.3cm}
\begin{center}
\begin{tabular}{ll}
Example I:  & $x_{i,t} = \phi_i x_{i,t-1} + \epsilon_t,~~ \text{for}~~ i=1,\cdots,p$, \vspace{-0.3cm}\\
Example II:  &  $x_{i,t} = \phi_i x_{i,t-1}   - \phi_i^{-1}\epsilon_{t-1} + \epsilon_t,~~ \text{for}~~ i=1,\cdots,p$, 
\end{tabular}
\end{center}\vspace{-0.2cm}

\noindent here $\{\phi_i\}_{i=1}^p$ are randomly generated from Uniform(0.5, 0.9). For Example I, $\epsilon_t$ is generated from i.i.d. standard normal, student's t distribution with 3 degrees of freedom, or exponential distribution with $\lambda=1$. For Example II, $\epsilon_t$ is generated from all distributions considered in Example I except for the standard normal so that each component $x_{i,t}$ is not independent across time. Similar to the simulation study in \cite{matteson2011dynamic}, we set $p=5, ~10$, $n=50, ~100, ~200$, and $h_0=1, ~2$, where the results for $h_0=2$ is reported in the supplementary material which shows similar results as the ones in Table~\ref{table:SMIC}. Also, the finite sample performance for larger dimension is summarized in the supplementary material. We note that $\X_t$ are mutually mean independent and fall into the framework (\ref{eq:single_model}). Example I with standard normal $\epsilon_t$ is adopted from \cite{matteson2011dynamic} and the data has linear dependence since all components are generated from linear time series models. For Example II, the data has no linear dependence but has a dependence beyond linear since $\X_t$ is generated from the all-pass ARMA(1,1) model which is a well known example for temporally uncorrelated but causal process; see \citet{lobato2002testing}.  
Therefore, the existing approaches that rely on the autocovariance may not be able to detect the important dynamics since the linear metric is not capable to measure the nonlinear dependence.  

We apply our approach and compare with the existing approaches, the dynamic orthogonal components (DOC) in \citet{matteson2011dynamic} and the second-order blind source separation (SOBI) in \citet{belouchrani1997blind}, which search for the uncorrelated components. In order to compare the accuracy of the estimation of $\A$, we compute the scaled version of the distance between two orthogonal matrices so that $\mathcal{D}^2(\A, \widehat{\A}) \in [0,1]$, i.e., $\mathcal{D}^2(\A, \widehat{\A}) = \left(1-\frac{1}{p}\sum_{i=1}^p  \left| \A_{\cdot i}^\top \widehat{\A}_{\cdot i}\right|\right) 
$ where smaller $\mathcal{D}^2$ indicates more accurate result. 

Table~\ref{table:SMIC} summarizes the results based on 500 replications. Overall, both existing methods and our approach produce more accurate dimension reduction results when $n$ increases. For Example I with a standard normal $\epsilon_t$, we observe that three approaches produce comparable results but the DOC and the SOBI methods perform slightly better than our method. Better perfomances from the DOC and the SOBI are expected since the data is generated from Gaussian linear time series and the entire dependence can be captured by the covariances. However, our approach outperforms the existing methods when the data is generated by a student's t distribution or an exponential distribution. For Example II, our approach is superior to the DOC and the SOBI methods for all sample sizes and different distributions. This demonstrates the robustness of our approach for the dependence and the distribution.

\subsection{Group Mean Independent Component Analysis}
In this section, we study the finite sample performance for the group mean independent component analysis (GMICA) in Section~\ref{sec:GMIC}. We consider two examples, where Example I is adopted from \citet{chang2018principal}. More specifically, the data is generated by $\Y_t = \A \X_t$, where $\A$ is defined in Section~\ref{subsec:sim_SMIC}, and $\X_t =\left\{ (\X_{t}^{(1)})^\top, (\X_{t}^{(2)})^\top, (\X_{t}^{(3)})^\top\right\}^\top=(x_{1,t}, \cdots, x_{p,t})^\top$, $m=3$, is generated by $x_{i,t} = z_{t+i-1}^{(1)}, ~ i\leq p/2$, $x_{i,t} = z_{t+i-p/2-1}^{(2)},~ p/2+1\leq i\leq p\cdot5/6$, $x_{i,t} =z_{t+i-p\cdot5/6-1}^{(3)},~ p\cdot5/6+1\leq i\leq p$,
\vspace{-0.3cm}
\begin{center}
\begin{tabular}{ll}
Example I:  
& $z_{t}^{(1)}=0.5z_{t-1}^{(1)}+0.3z_{t-2}^{(1)} + \epsilon_t^{(1)}- 0.9\epsilon_{t-1}^{(1)} +0.3\epsilon_{t-2}^{(1)}+1.2\epsilon_{t-3}^{(1)}+1.3\epsilon_{t-4}^{(1)}$, \vspace{-0.3cm}\\
&$z_{t}^{(2)}=0.8z_{t-1}^{(2)}-0.5z_{t-2}^{(2)} + \epsilon_t^{(2)}+\epsilon_{t-1}^{(2)}+0.8\epsilon_{t-2}^{(2)} +1.8\epsilon_{t-3}^{(2)}$, \vspace{-0.3cm}\\
&and $z_{t}^{(3)}=-0.7z_{t-1}^{(3)}-0.5z_{t-2}^{(3)} + \epsilon_t^{(3)}-\epsilon_{t-1}^{(3)}-0.8\epsilon_{t-2}^{(3)}$.\vspace{-0.3cm} \\
Example II:  
& $z_{t}^{(1)}=0.9 z_{t-1}^{(1)}   - 0.9^{-1}\epsilon_{t-1}^{(1)} + \epsilon_t^{(1)}$, $z_{t}^{(2)}=0.5 z_{t-1}^{(2)}   - 0.5^{-1}\epsilon_{t-1}^{(2)} + \epsilon_t^{(2)}$,\vspace{-0.3cm}\\
& and $z_{t}^{(3)}=0.1 z_{t-1}^{(3)}   - 0.1^{-1}\epsilon_{t-1}^{(3)} + \epsilon_t^{(3)}$.
\end{tabular}
\end{center}\vspace{-0.2cm}

\noindent Similar to examples in Section~\ref{subsec:sim_SMIC}, $\epsilon_t^{(1)},~ \epsilon_t^{(2)},~\epsilon_t^{(3)}$ are gernerated from standard normal, student's t distribution with 3 degrees of freedom, or exponential distribution with $\lambda=1$. 
We further note that $(\X_{t}^{(1)}, \X_{t}^{(2)}, \X_{t}^{(3)})$ are mutually uncorrelated and mean independent in Example I and each series $z_{t}^{(i)}, ~i=1,2,3$ has linear dependence. To check the performance for the dependence beyond linear dependence, we replace $(z_{t}^{(1)},z_{t}^{(2)},z_{t}^{(3)})$ in Example I with the series generated from all-pass ARMA(1,1) for Example II, where each series is uncorrelated; see \citet{lobato2002testing}. By following the simulation study in \citet{chang2018principal}, we set $p=6, ~12$, $n=200, ~500, ~1000$ and $h_0=1,~5$, where the results for $h_0=1$ is reported in the supplementary material and shows similar results for our approach. Also, the finite sample performance for larger dimension is summarized in the supplementary material.  

We compare the performance of our approach in comparison with the existing method, TS-PCA in \cite{chang2018principal}, which searches for the uncorrelated low-dimensional components. For the fair comparison, we treat the group information is unknown and estimate both the group structure and the separation matrix $\A$ with our objective function using Algorithm \ref{alg:group} with $c_0=0.75$ and no iterations since the method of \citet{chang2018principal} does not iteratively search for $\A$. In other words, we shall estimate the group structure and $\A$ through Step 1 - Step 6 in Algorithm \ref{alg:group}. By following \cite{chang2018principal}, we compute $\pi\in [0,1]$ the proportions of correct segmentations, and compute the scaled version of the $\widetilde{\mathcal{D}}^2$-distance when the method correctly group the components; see the supplemental material of \cite{chang2018principal}, i.e., $\pi = \frac{\sum_{l=1}^{R}{\bf 1}\left\{( \widehat{p}_{1, l}, \cdots, \widehat{p}_{\widehat{m}, l})=( p_1, \cdots, p_m)\right\} }{R}$, $\widetilde{\mathcal{D}}^2(\A, \widehat{\A}) =\max_{i=1, \cdots, m} \left(1-\frac{1}{p_i}  tr\left(\A_{i}\A_{i}^\top \widehat{\A}_{i}\widehat{\A}_{i}^\top \right)\right)$, 
where $R$ is the number of the replications, $\{\widehat{p}_{i, l}\}_{i=1}^{\widehat{m}}$ is the estimate of the number of components for each group for $l$th replicate. Also, notice that a higher $\pi$ and a smaller $\widetilde{\mathcal{D}}^2$ indicate more accurate result. 

Table~\ref{table:GMIC} presents the results based on 500 replications. From Table~\ref{table:GMIC}, we observe that our approach outperforms the TS-PCA for both Example I and Example II in terms of higher $\pi$ and smaller $\widetilde{\mathcal{D}}^2$. The difference between our approach and the TS-PCA becomes more prominent for Example II. In Example II, we consider latent components that are white noise but have nonlinear dependence. Therefore, the existing method with a covariance
matrix cannot detect the components. This could be a part of the reason for the larger $\widetilde{\mathcal{D}}^2$-distances produced by the existing approach. 

\begin{table}[!h]
	\caption{\small Reported are the average and the standard deviation of $\mathcal{D}^2$-distance with $h_0=1$
based on 500 replications. Three methods are compared: our method
(MICA), the method in \citet[DOC]{matteson2011dynamic}, and the method in \citet[SOBI]{belouchrani1997blind}.}
\label{table:SMIC} 	
	\label{tab:sim1}
	\vspace{0.3cm}
	\centering
\scalebox{.55}{
		\begin{tabular}{c c c c c c c}
\toprule
			&Dist.	& $p$ & Methods &  $n=50$ & $n=100$  & $n=200$\\  
\midrule
			\multirow{18}{*}{Example I}&\multirow{6}{*}{Normal}  & \multirow{3}{*}{5} 
			&MICA & 0.068 (0.027)  &   0.056 (0.027)    &   0.050 (0.028)                                     \\
			&&  &DOC & 0.064 (0.028)  &  0.048 (0.026)  &  0.036 (0.024)             \\
			&&  &SOBI &  0.062 (0.027)& 0.049 (0.026)   &    0.033 (0.023)          \\			
			\cmidrule{3-7} 
			&&\multirow{3}{*}{10}   &MICA &  0.149 (0.028)  &  0.137 (0.026) &   0.124 (0.026)               \\
			&&  & DOC & 0.143 (0.024)  &  0.125 (0.026) &  0.101 (0.027)                        \\
			&&  &SOBI &  0.145 (0.024) & 0.123 (0.027)  &      0.101 (0.026)        \\					
			\cmidrule{2-7} 
			&\multirow{ 6}{*}{ $t$}  & \multirow{ 3}{*}{5} 
			&MICA    & 0.050 (0.033)   & 0.026 (0.023)  & 0.012 (0.014)                          \\
			&&  &DOC  &  0.078 (0.041)  &   0.062 (0.035)    & 0.045 (0.039)                                                   \\
				&&  &SOBI & 0.085 (0.044) & 0.061 (0.038)   &   0.048 (0.041)           \\				
				\cmidrule{3-7} 
			&&\multirow{ 3}{*}{10}   &MICA & 0.123 (0.038)   & 0.072 (0.028)  &    0.034 (0.023)               \\
			&&  &DOC &    0.174 (0.046)    &   0.148 (0.047)   &  0.117 (0.042)                                                              \\
			&&  &SOBI & 0.180 (0.045)  &  0.156 (0.052) &      0.124 (0.044)        \\					
			\cmidrule{2-7} 
			&\multirow{ 6}{*}{exp}  & \multirow{ 3}{*}{5} 
			&MICA & 0.049 (0.026)   &  0.026 (0.022)   &  0.009 (0.013)                                  \\
			&&  &DOC &  0.085 (0.028) &    0.080 (0.029)       &  {0.061 (0.028)}                                           \\
			&&  &SOBI &  0.067 (0.030)&  0.050 (0.028)  &    0.035 (0.024)          \\					
			\cmidrule{3-7} 
			&&\multirow{ 3}{*}{10}   &MICA &  0.128 (0.026)&  0.092 (0.027)   & 0.047 (0.023)         \\
			&&  & DOC &   0.156 (0.027)   & 0.142 (0.026)    &  0.123 (0.027)                                                     \\
			&&  &SOBI &  0.159 (0.023) & 0.148 (0.027)  &      0.100 (0.026)        \\					
\midrule	
      \multirow{12}{*}{Example II}&\multirow{6}{*}{ $t$}  & \multirow{ 3}{*}{5} 
			&MICA & 	0.058 (0.037)&0.029  (0.025)  &   0.014 (0.015)\\
			&&  &DOC & 0.133 (0.064) &  0.123 (0.057) & 0.114 (0.056)\\
			&&  &SOBI &  0.094 (0.033)& 0.092 (0.032)  &       0.098 (0.031)       \\					
\cmidrule{3-7} 
&&\multirow{3}{*}{10}   &MICA &  0.136 (0.042)  & 0.081 (0.037)  & 0.039 (0.026)\\
&&  &DOC & 0.231 (0.058) & 0.224 (0.062) & 0.221 (0.062) \\
			&&  &SOBI & 0.184 (0.025) &  0.181 (0.025)  &    0.187 (0.026)          \\		
			\cmidrule{2-7} 
&\multirow{ 6}{*}{exp }  & \multirow{ 3}{*}{5} 
&MICA & 0.063 (0.034) &  0.027 (0.022) &  0.007 (0.009)\\
&&  &DOC & 0.112 (0.045) &  0.116 (0.050)  &  0.109 (0.044)\\
			&&  &SOBI & 0.118 (0.054)  & 0.119 (0.047)  &      0.113 (0.048)        \\		
\cmidrule{3-7}
&&\multirow{ 3}{*}{10}   &MICA &0.162 (0.037)&  0.113 (0.034) &  0.051 (0.023) \\
&&  & DOC &   0.212 (0.042)& 0.209 (0.040)  & 0.201 (0.039)\\
			&&  &SOBI & 0.214 (0.038)  &  0.215 (0.039) &    0.213 (0.038)          \\		
\midrule
		\end{tabular}
}
\end{table}

\begin{table}[!h]
	\caption {\small Reported are the proportions of correct segmentations, $\pi$, the average and the standard deviation of $\widetilde{\mathcal{D}}^2$-distance when estimated groups are correct with $h_0=5$ based on 500 replications. Two methods are compared: our method
(GMICA), the method in \citet[TS-PCA]{chang2018principal}. }
\label{table:GMIC} 
	\vspace{0.3cm}
	\centering
	\scalebox{.65}{
		\begin{tabular}{c c c c c c c c c c }
			\toprule
		 &	\multirow{2}{*}{Dist.} & \multirow{2}{*}{$p$} &\multirow{2}{*}{Method}  &  \multicolumn{2}{c}{$n=200$} &   \multicolumn{2}{c}{$n=500$}  &  \multicolumn{2}{c}{$n=1000$}    \\\cmidrule{5-6} \cmidrule{7-8}\cmidrule{9-10} 
			&  & & & ${\pi}$ & $\widetilde{\mathcal{D}}^2$ &  ${\pi}$ & $\widetilde{\mathcal{D}}^2$   &  ${\pi}$ & $\widetilde{\mathcal{D}}^2$   \\  \midrule %
			\multirow{12}{*}{Example I} &	\multirow{4}{*}{Normal} & \multirow{2}{*}{6}
			&GMICA &  0.684    & 0.029 (0.019)      &  0.884  & 0.012 (0.008)&  0.976 & 0.005 (0.003)\\
	    &	&	& TS-PCA & 0.682       & 0.101 (0.054)     &  0.886 &  0.048 (0.033)& 0.970  & 0.024 (0.016) \\
	    		\cmidrule{3-10} 
		&	 & \multirow{2}{*}{12}
            & GMICA&  0.364    & 0.065 (0.069)      &  0.756 & 0.021 (0.017) &  0.882 & 0.017 (0.011) \\	    	
        &  &	& TS-PCA &   0.152    &  0.216 (0.061)  & 0.288  &0.149 (0.056)  & 0.460 & 0.111 (0.048)\\
			\cmidrule{2-10}				
		&	\multirow{4}{*}{$t$} & \multirow{2}{*}{6}
			& GMICA &  0.638   & 0.022 (0.022)      &  0.858  & 0.007 (0.006)& 0.974 & 0.004 (0.003)\\
		&	&	& TS-PCA & 0.612       & 0.120 (0.069)     &  0.842 &  0.057 (0.047)& 0.938  & 0.031 (0.027) \\
			\cmidrule{3-10} 
		&	& \multirow{2}{*}{12}
			&GMICA &  0.414   & 0.050 (0.022)      &   0.714& 0.031 (0.050) &  0.816 & 0.015 (0.010) \\	 
	&		&	& TS-PCA &   0.164    & 0.216 (0.076)     &   0.372  & 0.156 (0.055) & 0.462 & 0.120 (0.057)\\
			\cmidrule{2-10}				
	&		\multirow{4}{*}{$\exp$} & \multirow{2}{*}{6}
			&GMICA &  0.640   & 0.023 (0.028)     &  0.880  & 0.008 (0.005)&  0.956 & 0.004 (0.003)\\
	&		&	& TS-PCA & 0.630       & 0.104 (0.055)     &  0.842 &  0.045 (0.026)& 0.938  & 0.023 (0.015) \\
			\cmidrule{3-10} 
	&		& \multirow{2}{*}{12}
			& GMICA &  0.454   & 0.064 (0.043)      &   0.752 & 0.030 (0.031) &  0.790 & 0.016 (0.008)\\	 
	&		&	& TS-PCA &   0.138    &  0.218 (0.061)   &    0.338 & 0.149 (0.049)  & 0.426& 0.108 (0.046) \\
			\hline							
	\multirow{8}{*}{Example II} &	\multirow{4}{*}{$t$} & \multirow{2}{*}{6}
& GMICA &  0.608  & 0.069 (0.084)      &  0.702 & 0.039 (0.057)&  0.728 & 0.023 (0.034)\\
&	&	& TS-PCA & 0.158       & 0.229 (0.062)     &  0.160  & 0.198 (0.071)  & 0.132 & 0.171 (0.066) \\
\cmidrule{3-10} 
&	& \multirow{2}{*}{12}
& GMICA &  0.306   & 0.075 (0.043)      & 0.468  & 0.042 (0.033)& 0.660  & 0.036 (0.029) \\	 
&		&	& TS-PCA &     0.098  &   0.247 (0.058)  &   0.088 & 0.214 (0.054) & 0.096  &   0.194 (0.059) \\
			\cmidrule{2-10}				
&		\multirow{4}{*}{$\exp$} & \multirow{2}{*}{6}&GMICA &  0.768  & 0.067 (0.062)      &  0.920  & 0.028 (0.029)& 1.000 & 0.012 (0.010)\\

&		&	& TS-PCA & 0.136   &  0.234 (0.064)   &    0.088 &0.214 (0.056)  &  0.064& 0.200 (0.056)  \\
\cmidrule{3-10} 
&		& \multirow{2}{*}{12}
& GMICA&  0.354   & 0.098 (0.046)      &  0.556 & 0.032 (0.022) &  0.808 & 0.024 (0.019) \\	 
&		&	& TS-PCA &   0.086    &   0.267 (0.048)  &   0.044  & 0.234 (0.053) &  0.050    &   0.222 (0.057) \\
\midrule								   			
		\end{tabular}
	}
\end{table}

\section{Real Data Illustriation}\label{sec:real data}

In this section, we illustrate our approach by forecasting Google's community mobility reports during COVID-19. The data contains daily changes in mobility trends compared to a baseline established on February 15, 2020. It captures movements across 6 categories of locations. 
It is essential to produce accurate predictions on the community mobility as it is related to a virus spread and support to combat the virus.

In our analysis, we consider community mobility reports of eight states in Northeastern of the U.S., including Connecticut, Maine, Massachusetts, New Hampshire, Rhode Island, Vermont, New York, and New Jersey. By following \cite{yaoming2023non}, we also consider the average of the mobility of all 6 categories in the particular day and state from January 1, 2021 to February 28, 2022, thus we have $\Y_t \in \R^8$ of length $n=424$. In order to make the data stationary, we first detrend each time series by using polynomial smoothing to remove the trend and standardize the time series with $\var(\Y_t)^{-1/2}$ so that the data falls into the framework (\ref{eq:single_model}) or (\ref{eq:group_model}). The auto and cross correlations of $\Y_t$ which is reported in the supplementary material exhibits weekly periodicity. 
Hence we set $h_0=7, ~14$ and apply both MICA and GMICA. We first divide the data into a training set and a testing set, where the data from January 1, 2021 to December 31, 2021 (365 data points) is assigned as a training set and the remaining two months of observations starting from January 1, 2022 to February 28, 2022 (59 data points) are assigned as a testing set. To compare the prediction accuracy,  we compute the mean squared prediction error (MSPE) where smaller MSPE indicates more precise prediction, i.e., $\text{MSPE} = \frac{\|\widehat{\Y}^{(q)}_{n_0+j+q} - \Y_{n_0+j+q}\|^2}{p},$ where $\widehat{\Y}^{(q)}_{n_0+j+1}$ is the $q$-step ahead forecasting of $ \Y_{n_0+j+q}$, where $n_0=365$, $j=1,\cdots, 58$, and $q=1,2$.
To produce $\widehat{\Y}^{(q)}_{n_0+j+1}$, we follow the approach in \citet{chang2018principal} which is described by the following steps. (1) We first apply the proposed MICA and GMICA to the training set as described in our simulations, and obtain $\widehat{\A}$ and $(\widehat{p}_1, \cdots, \widehat{p}_{\widehat{m}})$ which are fixed afterward. (2) We estimate $\{\widehat{\X}_t\}_{t=1}^{n_0+j}$ by $\{\widehat{\A}^\top \Y_t\}_{t=1}^{n_0+j}$. (3) Due to the existence of the seasonal correlations, we fit appropriate univariate or multivariate seasonal ARMA $(d_1, 0) \times (d_2, 0)_7$ models to each component or group of $\widehat{\X}_t$, and obtain $q$-step ahead predictions of the components, $\widehat{\X}^{(q)}_{n_0+j+1}$, where the orders $d_1$ and $d_2$ are determined by AIC. (4) We finally generate the prediction for ${\Y}_{n_0+j+q}$ by $\widehat{\Y}^{(q)}_{n_0+j+q} = \widehat{\A} \widehat{\X}^{(q)}_{n_0+j+q}$. In addition to our MICA and GMICA, we apply 6 different approaches to compare the prediction accuracy. In particular, we further apply the existing methods TS-PCA of \cite{chang2018principal}, DOC of \citet{matteson2011dynamic}, SOBI of \citet{belouchrani1997blind} to this data, where the predictions are generated by following the same procedure above. Also, we consider the parametric models by fitting univariate seasonal ARMA model  (sUARMA) for each components in $\Y_t$ and fitting seasonal vector ARMA model (sVARMA) to $\Y_t$ by following step (3). Lastly, we consider fitting a refined VAR model (refVAR) to the data $\Y_t$ with a thresholding value equals to 1, and the order selected by AIC. 

\begin{table}[H]
\caption{\small Avearage MSPE for community mobility data.}
	\centering
	\scalebox{0.55}{\begin{tabular}{c c c c }
			\hline		Method	&&  $h_0=7$ &$h_0=14$ \\ 
			\hline
			\multirow{3}{*}{GMICA}  &  Groups &  6 Groups (3,1,1,1,1,1)&6 Groups (3,1,1,1,1,1)   \\
			& MSPE ($q=1$) &  1.003 &1.002   \\
			& MSPE ($q=2$) &   0.960 &0.952  \\		
\hline
			\multirow{3}{*}{MICA}  &  Groups  & 8 Groups & 8 Groups   \\
			& MSPE ($q=1$)& 1.011 & 1.023  \\
			& MSPE ($q=2$) &   0.993 & 0.989\\		    
					\hline
			\multirow{3}{*}{TS-PCA}  &  Groups &  7 Groups (2,1,1,1,1,1,1)  & 7 Groups (2,1,1,1,1,1,1)    \\
			& MSPE ($q=1$) &   1.112 &1.139   \\
			& MSPE ($q=2$) &   1.164 &1.041   \\	
\hline
 \multirow{3}{*}{DOC}  &  Groups  & 8 Groups & 8 Groups   \\
			& MSPE ($q=1$) & 1.127 & 1.123  \\
			& MSPE ($q=2$) &   1.042 & 1.064\\	
\hline
\multirow{3}{*}{SOBI}  &  Groups  & 8 Groups & 8 Groups   \\
			& MSPE ($q=1$) & 1.122 & 1.155  \\
			& MSPE ($q=2$) &  1.036 & 1.045\\	
			\hline
			\multirow{3}{*}{sUARMA}  &  Groups &  8 Groups & 8 Groups  \\
			& MSPE ($q=1$)  &  1.235 &1.235 \\
			& MSPE ($q=2$) &   1.281  & 1.281  \\		 
			\hline
			\multirow{3}{*}{sVARMA}  &  Groups &1 Group & 1 Group   \\
			& MSPE ($q=1$) &  1.159 &  1.159 \\
			& MSPE ($q=2$) & 1.231 & 1.231 \\		 
			\hline
			\multirow{3}{*}{refVAR}  &  Groups  & 1 Group &1 Group      \\
			& MSPE ($q=1$) & 1.583 & 1.583 \\
		 & MSPE ($q=2$) &  2.353&  2.353 \\		 
			\hline
	\end{tabular}}
		\label{msfe}
\end{table}

Table \ref{msfe} summarizes the prediction accuracy for all methods. From Table \ref{msfe}, our GMICA estimates 6 different groups, where the largest group has 3 components. On the other hand, TS-PCA generates 7 different groups. Thus, the number of groups are different for TS-PCA and our method, where the number of groups for TS-PCA is larger. This aligns with our argument in Remark~\ref{remark:1}. 
In terms of the predictions, our GMICA generates notably smaller MSPE than the ones of the existing methods including TS-PCA, DOC, SOBI, and parametric models sUARMA, sVARMA, and refVAR. In particular, MSPE of GMICA is roughly 10\% lower than MSPE of TS-PCA, DOC, SOBI on average for $q=1$. Also, the difference of MSPE becomes more prominent by comparing GMICA and parametric models sUARMA, sVARMA, and refVAR. Moreover, GMICA performs better than MICA. We believe this is due to the extra grouping step in the GMICA and forming groups for the components seems to be helpful for this data.  Notice that GMICA forms groups if the past or current values of another components affect the conditional mean of the other components.  Thus, if the main interest is the prediction, it would be more beneficial to seek the group mean independent components and build separate time series model for each group since GMICA loses least amount of loss on the conditional mean which is the optimal predictor in terms of the mean squared error sense. We conjecture that this could be one reason for better prediction performance of GIMCA. It is interesting to observe that our MICA still produces more accurate predictions than the other existing methods such as TS-PCA, DOC, SOBI, sUARMA, sVARMA, and refVAR in terms of smaller MSPE. We note that sUARMA, sVARMA, and refVAR produce the same prediction performance under $h_0=7$ and $h_0=14$ since those approaches do not depend on $h_0$. It is worth mentioning that the improvement in forecasting for our approaches GMICA, MICA compared to the existing methods TS-PCA, DOC, SOBI is essentially due to the different estimator of $\A$, since we apply the same procedure to generate the predictions. For this data, our approach effectively separates the components that lead to more accurate forecastings.  We additionally compare the prediction performances focusing on each category which are reported in our supplementary material. We observe that GMICA or MICA consistently produce smaller MSPE across all categories. 

We further report the auto and cross correlations of $\widehat{\X}_t$ for our GMICA and the existing approach TS-PCA in our supplementary material in order to check the uncorrelatedness across different groups.  By comparing Figures S2 and  Figures S3, it appears that our GMICA shows weaker linear dependence between two components that belong to different groups. For instance, the third component of TS-PCA seems to be correlated with the first and second components which are in a separate group. This could be a part of the reason that our GMICA can produce more accurate forecastings for this data. We observe a similar phenomenon in Figure~S2 from the auto and cross correlations of $\widehat{\X}_t$ for our MICA and the DOC and the SOBI which are reported in the supplementary material.

\section{Summary And Conclusions}\label{sec:conclusion}
We propose new approaches to model the multivariate time series data by seeking the latent mean independent components. Once the mean independent components are identified, a separate analysis or model can be built for each component. This reduces huge number of parameters involved in the modeling. Compared to the existing methods, our approach can be more robust and flexible to different forms of dependence and different distributions of the data, which is demonstrated in our simulation study. We further introduce a natural extension to the group mean independent component analysis by following the idea of \citet{chang2018principal}. To justify the validity of our method, we provide the asymptotic properties under two scenarios: fixed $p$ and growing $p$ with the sample size $n$. In addition to encouraging finite sample performance, we further illustrate the advantage of our approaches with a real data application that can generate more accurate predictions.

We conclude this article by discussing several future research topics. We introduce a procedure to estimate the group structure. 
Our simulation results indicate that the method identifying the group structure produces reasonable results, however there is no theoretical justification. It would be helpful to provide the theoretical property for the estimation of the group structure. Also, extending our group mean independent component analysis to allow $m\rightarrow \infty$ would be useful. Those would require new theoretical investigations. 
Furthermore, imposing the sparsity assumption on $\mathbf{A}$ for the group mean independent component analysis is an interesting direction and we conjecture that the sparse method may allow $p$ to grow faster than $p/\sqrt{n} \rightarrow 0$ as $n\rightarrow \infty$. It requires an updated objective function, along with a different estimation procedure. 
Lastly, we often observe matrix or tensor time series data, and the dimension reduction for matrix or tensor time series has received considerable attention; see \citet{wang2019factor}, \citet{chen2019constrained}, \citet{chen2021factor} for the recent development for the dimension reduction on matrix or tensor time series. It would be useful to extend our approach for matrix or tensor time series to reduce the number of the parameters. The research along these directions are well underway.

\section*{Supplementary Material} 

Supplementary material available online includes technical proofs of theoretical results and state additional theorem with its proof, and reports additional simulations, real data application results, and figures.

\section*{Acknowledgements}
The authors thank the Editor, the Associate Editor, and two referees for their constructive comments and suggestions that led to substantial improvements. 

\bibliographystyle{agsm}

\bibliography{MICA_ref}

\end{document}